\documentclass[10pt]{article}

\usepackage[dvips]{graphicx}
\usepackage{amssymb}
\newcommand{\etal} {\eta_{_L}}
\newcommand{\etas} {\eta_{_S}}
\newcommand{\etap} {\eta_{_P}}

\twocolumn
\begin{document}

\noindent
{\bf Damping of Acoustic Waves in Dilute\\ Polymer Solutions}\\~\\
{\bf Y. Tsori and P.-G. de Gennes}\\
{\it Physique de la Mati\`{e}re Condens\'{e}e, Coll\`{e}ge de\\
France,
Paris, France}\\




{\bf 1. Introduction.}
The shear viscosity of dilute polymer solutions $\etas(\omega)$
has been studied in the past in experiments and models and
analyzed in a classic book by Ferry.$^{\cite{ferry}}$ The
standard view is that each polymer coil has a sequence of 
modes discussed by Zimm$^{\cite{zimm}}$ which are due to
hydrodynamic interactions between the monomers; the viscosity
$\etas$ then appears as a weighted sum of Lorentzian responses
from these modes. For qualitative purposes, at frequencies which
are not too high, the first mode dominates, as noticed very early
by Peterlin,
and $\etas$ can be reduced to a simple form,
\begin{eqnarray}\label{shear}
\etas(\omega)=\alpha_s\nu k_BT\frac{\tau_z}{1+w^2\tau_z^2}
\end{eqnarray}
Here $\alpha_s$ is a coefficient of order unity, $\nu$ is the
number of polymer coils per unit volume, $k_BT$ is the thermal
energy, $\omega$ is the angular frequency, and $\tau_z$ is the
Zimm relaxation time, with the scaling structure$^{\cite{pgg}}$
\begin{eqnarray}
\tau_z\simeq \frac{\eta  R_F^3}{k_BT}
\end{eqnarray}
Here $\eta$ is the solvent shear viscosity, and $R_F$ is the coil
size, which we will estimate for good solvents following Flory's
approach$^{\cite{pgg,flory}}$
\begin{eqnarray}\label{flory}
R_F^5=a^5\left(1-2\chi\right)N^3
\end{eqnarray}
where $a$ is the monomer size, $N$ is the number of monomers in a
chain, and $\chi$ is the Flory interaction
parameter.$^{\cite{flory}}$

Our aim in the present note is to discuss the longitudinal
viscosity $\etal(\omega)$ which may be measured via attenuation of
ultrasound = this implies that we now consider the solution as a
(weakly) compressible fluid. For incompressible fluids, symmetry
imposes that $\etal=\frac43\etas$. But here we must write
\begin{eqnarray}
\etal=\frac43\eta_s+\eta_p
\end{eqnarray}
where $\etap$ includes specific effects of density changes, or
(equivalently) of the applied pressure $p$ (see section 3).

Our basic idea is that $p$ modifies the $\chi$ parameter, and thus
modifies $R_F$, as indicated in eq \ref{flory}.  We construct a
crude model for the resulting dynamics, assuming an analogue of
Peterlin's approach, with a single mode relaxation time $\tau_z$.

The equilibrium results are presented in section 2, and the
dynamics is schematized in section 3. The surprising conclusion is
that $\etap$ is at least comparable to $\etas$, even in liquids
which are not very compressible; this is discussed in section 4.
All the exploration is made only at the level of scaling laws.


{\bf 2. Basic Couplings Between Coil Size and Pressure.}
We start from a coil in good solvent, with a size described by the
Flory expression, eq \ref{flory}. We then raise the pressure by an
amount $p$; the equilibrium size shifts by a certain amount
$\delta R$. We can think of (at least) two contributions to
$\delta R$.

The first contribution, $\delta R_K$, is due to a change in Kuhn
length, which may be written in the form
\begin{eqnarray}
\frac{\delta R_K}{R_K}=\alpha_K\frac{p}{K}
\end{eqnarray}
where $K\simeq 10^3$ MPa is the elastic modulus of the solvent,
$p/K$ measures the solvent contraction, and $\alpha_K$ is a number
which is expected to be of order unity (and which may be positive
or negative).

The second contribution, $\delta R\chi$, is the consequence of a
change $\delta\chi$ in the Flory parameter $\chi$,
\begin{eqnarray}
\delta\chi=\alpha_\chi \frac{p}{K}
\end{eqnarray}
Again we expect $|\alpha_\chi|$ to be of order
unity.$^{\cite{chi_p1,chi_p2}}$ The resulting shift in size is
obtained by differentiating eq \ref{flory},
\begin{eqnarray}\label{deltaRchi}
\frac{\delta R_\chi}{R_F}=-\frac25\frac{\delta\chi}{1-2\chi}
\end{eqnarray}
Equations \ref{flory} and \ref{deltaRchi} hold provided that we
are in a strong swelling regime, i.e., that the perturbation
parameter $\zeta$ describing the excluded volume is 
small.$^{\cite{yamakawa}}$ The scaling structure of
$\zeta$ is $\zeta=N^{1/2}(1-2\chi)$ and thus eq \ref{deltaRchi}
assumes that
\begin{eqnarray}
1-2\chi>N^{-1/2}
\end{eqnarray}

We arrive now to an interesting point. When the pressure $p$ is
applied, the solvent is contracted = all linear dimensions are
reduced by a factor $1-p/3K$. If the relative shift of $\delta R$
was just equal to $-p/3K$, the whole system would be just modified
by an affine deformation, and no relaxation process would be
involved. Thus we must define an effective chain deformation
$\delta R_{\rm eff}$, where the affine contribution has been taken
out. Collecting all these results we arrive at
\begin{eqnarray}
\frac{\delta R_{\rm eff}}{R_F}&=&\alpha\frac{p}{K}\\
\alpha&=&\alpha_K-\frac25\frac{\alpha_\chi}{1-2\chi}-\frac13\label{alpha}
\end{eqnarray}
However, we immediately note that the $\alpha_\chi$ term will
dominate if
\begin{eqnarray}
1\gg 1-2\chi\gg N^{-1/2}
\end{eqnarray}
%


{\bf 3. The Longitudinal Viscosity.}
For a compression wave in a liquid, individual molecules can move
parallel or perpendicular to the direction of wave propagation.
The existence of these two modes means that energy dissipation
depends on two independent quantities, hence there are two
fundamental viscosities. The longitudinal compression mode brings
about a term ${\bf div}\cdot{\bf v}$ in the Navier--Stokes
equation, a term which is absent in the more usual case of an
incompressible liquid. In the linear theory employed here, it is
sufficient to consider only monochromatic waves; the superposition
of many modes is trivial. For such a longitudinal wave in the
$x$-direction, the velocity of molecules is given by
\begin{eqnarray}
v=v_x=v_0\cos(kx+wt)
\end{eqnarray}
where $\omega$ is the frequency, $c=\omega/k\approx 1500$ m/s is
the wave velocity ($c\gg v_0$), and $k$ is the wavevector. For
frequencies up to megahertz, the sound wavelength is much larger
than the coil size. Each chain then feels a time-varying pressure
$p=p_0 e^{i\omega t}$, with $v_0/c=p_0/K$. It can be shown$^{\cite{LL}}$ 
that the energy dissipation per unit volume is
given by
\begin{eqnarray}
T\dot{S}&=&\frac12 v_0^2k^2\left(\frac43\etas+\etap\right)\\
&=&\frac12 \left(\frac{\omega
p_0}{K}\right)^2\etal(\omega)\label{LL_exp}
\end{eqnarray}
where we have defined
$\etal(\omega)\equiv\frac43\etas(\omega)+\etap(\omega)$. $\etap$
is comparable to $\etas$ and always present in a longitudinal
wave. However, in the presence of internal slow relaxation
processes (e.g., slow chemical reactions) it is enhanced. In the
following, our aim is to find the expression for $\etal$ relevant
to polymer solutions.

It is important to realize that, for an acoustic wave, the
coefficient $\alpha$ is renormalized by a thermal effect.
The wave is adiabatic with temperature modulation
\begin{eqnarray}
\delta T=qp
\end{eqnarray}
where $q=(T/C_pV)(\partial V/\partial T)_p$ and $V$ is the volume.

For instance, the Kuhn length term $\alpha_K$ is renormalized to a
new value
\begin{eqnarray}
\tilde{\alpha}_K=\alpha_K+\frac{q}{R_K}\frac{dR_K}{dT}
\end{eqnarray}
We expect a similar renormalization for $\alpha_\chi$.
This renormalization should not affect the scaling law which we
postulated in section 2.

As we saw in the previous section, the equilibrium size of the
coil is modified from its Flory radius $R_F$ by an amount $\alpha
pR_F/K$ proportional to the external pressure. The polymer,
characterized by a ``response time'' $\tau_z$, does not follow
this size instantaneously, but rather with a certain time lag.
This time lag leads to dissipation. Thus we may denote $\delta R$
as the instantaneous deviation of the coil size from the Flory
size, and write
\begin{eqnarray}
\delta\dot{R}=\frac{1}{\tau_z}\left(\alpha\frac{p}{K}R_F-\delta
R\right)
\end{eqnarray}
The solution is straightforward: $\delta R=\delta R_0 e^{i\omega
t}$ and
\begin{eqnarray}
\delta R_0=\frac{\alpha}{K}\frac{p_0}{1+i\omega \tau_z}R_F
\end{eqnarray}

An equivalent formulation is to imagine an oscillating spring in a
viscous media characterized by a friction constant $\xi$,
\begin{eqnarray}
\xi\delta\dot{R}=\frac{k_BT}{R_0^2}\left(\alpha\frac{p}{K}R_F-\delta
R\right)
\end{eqnarray}
Here the spring constant is $k_BT/R_0^2$. Hence, we identify the
friction constant as $\xi=\tau_z k_BT/R_0^2$, and the total
dissipation per unit volume is then
\begin{eqnarray}
T\dot{S}&=&\frac12\nu \xi |\delta\dot{R}|^2\nonumber\\
&=&\frac12\nu\tau_z\frac{k_BT}{R_0^2}\left(\frac{\alpha
p_0}{K}\right)^2R_F^2\frac{\omega^2} 
{1+\omega^2\tau_z^2}\nonumber\\
&\simeq&\frac12\nu\tau_z k_BT\left(\frac{\omega p_0}{K}\right)^2
\frac{1}{(1-2\chi)^2}\frac{1}{1+\omega^2\tau_z^2}
\end{eqnarray}
In the last line we assumed that $\alpha$ is dominated by the
$\alpha_\chi$ term (eq \ref{alpha}), and that $\alpha_\chi$ is of
order unity.

Comparison with eq \ref{LL_exp} leads to the identification of
$\etal$ as
\begin{eqnarray}
\etal(\omega)=\nu\tau_z k_BT
\frac{1}{(1-2\chi)^2}\frac{1}{1+\omega^2\tau_z^2}
\end{eqnarray}
This expression is identical in form to the one-mode expression
for the shear viscosity $\etas$ in polymer solutions (eq
\ref{shear}), {\it except} for the $(1-2\chi)^{-2}$ prefactor,
which can be large in the vicinity of the $\Theta$ point.

How is the wave damped? In the above we have assumed that the
wavevector $k$ is purely real. In fact, damping is described by an
imaginary part of $k$ given in
\begin{eqnarray}
k=\frac{\omega}{c}+i\frac{1}{2\rho c^3}\etal\omega^2
\end{eqnarray}
with the imaginary part much smaller than the real one.


{\bf 4. Conclusions.}
(1) We expect a specific (pressure induced) longitudinal
viscosity $\etal(\omega)$ in polymer solutions. The pressure
contribution should be most visible near the $\Theta$ point, with
an enhancement factor
\begin{eqnarray}
f=\left(\frac{1}{1-2\chi}\right)^2\sim
\left(\frac{\Theta}{T-\Theta}\right)^2
\end{eqnarray}
(provided that $f$ remains smaller than $N$).
If we go below the $\Theta$ point, the pressure wave modulates a phase 
separation = there is another dynamics to consider (this was pointed out to us by 
a referee).
(2) We discussed the frequency dependence of $\etap$ in a
``one-mode approximation''. For the shear viscosity, the
restriction to one mode is acceptable when $\omega\tau_z<1$. But
the one-mode approximation could be worse for our case. We can
look at the related problem of chain collapse under a temperature
jump (from above $\Theta$ to below it). The early stages of
collapse involve ``local clumps'' along the chain = this is rather
different from the standard deformation in shear.
(3) Our main conclusion is that $\etap(\omega)$ may be
observable in certain dilute polymer solutions. Can we expect this
idea to hold for polymer melts? In a melt, the coils are
ideal (as first understood by Flory$^{\cite{flory}}$) and the
$\chi$ parameter becomes irrelevant. But we may still have a small
effect of the pressure on the Kuhn length.
(4) The pressure dependence of $\chi$ may also be of interest in
block copolymer melts. Consider, for instance, a lamellar
phase = a change in $\chi$ leads to a change in the interfacial
area per chain and to chain extension. However, in the strong
segregation limit the lamellar period scales as $\sim
N^{2/3}\chi^{1/6}$ and depends only weakly on the $\chi$ parameter.

\end{document}